\documentclass[conference]{IEEEtran}
\IEEEoverridecommandlockouts

\usepackage{cite}
\usepackage{amsmath,amssymb,amsfonts}
\usepackage{algorithmic}
\usepackage{graphicx}
\usepackage{textcomp}
\usepackage{xcolor}
\usepackage{flushend}
\usepackage[numbers]{natbib}
\def\BibTeX{{\rm B\kern-.05em{\sc i\kern-.025em b}\kern-.08em
    T\kern-.1667em\lower.7ex\hbox{E}\kern-.125emX}}
\pdfoutput=1
\begin{document}

\title{SoK: Evaluating Privacy and Security Concerns of Using Web Services for the Disabled Population}

\author{\IEEEauthorblockN{Alisa Zezulak,
Faiza Tazi, and
Sanchari Das}
\IEEEauthorblockA{
InSpirit Lab, University of Denver, Colorado; 
Emails: \{Alisa.Zezulak, Faiza.Tazi, Sanchari.Das\}@du.edu}}

\maketitle

\begin{abstract}

The online privacy and security of the disabled community is a complex field that has implications for every user who navigates web services. While many disciplines have separately researched the disabled population and their online privacy and security concerns, the overlap between the two is very high but under-researched. Moreover, a complex relationship exists between the disabled population and web services where the interaction depends on several web service developmental factors, including usability and accessibility. To this aid, we explored this intersection of privacy and security of web services as perceived by the disabled community through previous studies by conducting a detailed systematic literature review and
analysis of $63$ articles. Our findings encompassed several topics, including how the disabled population navigates around authentication interfaces, online privacy concerns, universal design practices, and how security methods such as CAPTCHAs can be improved to become more accessible and usable for people of all needs and abilities. We further discuss the gap in the current research, including solutions such as the universal implementation of inclusive privacy and security tools and protocols.

\end{abstract}

\section{Introduction}

The Covid-19 pandemic has necessitated people worldwide to adapt to new ways of doing things~\cite{beaunoyer2020covid}. With billions of people forced to conduct their daily activities online, including attending school, working from home, grocery shopping, banking, and other critical tasks~\cite{daniel2020education,aristovnik2020impacts,tazi2021parents,reddington2022development,monroe2021location,karmakar2021understanding}, the move to a fully digital world has been an inconvenience for some. Unfortunately, this drastic shift to online services has left many behind, particularly those who rely on usable, accessible, and inclusive services~\cite{scanlan2021reassessing,das2019security,das2020humans,das2019towards,das2018qualitative}. While the vulnerabilities of the disabled population have always existed, this sudden move to digital services has exacerbated existing problems~\cite{scanlan2021reassessing,bayor2018characterizing}, including privacy and security since vulnerable populations cannot use privacy and security tools and protocols successfully due to the disparities in usability and accessibility levels. Furthermore, these tools often fail to meet the specific requirements of the disabled population, even in fundamental areas such as authentication techniques~\cite{furnell2021disadvantaged,das2018johnny,das2019towards}.

Along with the usability and accessibility concerns, there are many data security and privacy concerns present, such as critical data access, smart home technology data usage, and inadequate authentication protocols. Additionally, the disabled population uses medical technology more than their non-disabled counterparts, but many of these tools and protocols are not accessible to users with different needs and abilities~\cite{liang2017understanding}. This makes accessing personal health records, and user accounts difficult for many users. Furthermore, the disabled population faces many difficulties online relating to authentication methods such as CAPTCHAs~\cite{helkala2012disabilities,das2020smart}. Most CAPTCHAs require a user to enter an alphanumeric code, which can be difficult or impossible for visually impaired users. This raises questions about if privacy and security tools are designed with different user populations in mind.

To provide a comprehensive understanding of the research undertaken in this area, we conducted a systematic literature review of $2,352$ research articles on the privacy and security of web services and the disabled populations. We screened these articles by title, abstract, and full text, selecting $63$ papers that focused on the privacy and security of web services as they relate to the disabled population. We then conducted a detailed thematic analysis of these papers, uncovering valuable solutions to address some privacy and security concerns of the disabled population. However, our analysis also revealed significant gaps in the research, highlighting the need for future work in this area. As far as we know, this is the first Systematization of Knowledge (SoK) paper to focus on the privacy and security challenges faced by the disabled community while accessing web services.

\section{Related Work}
\label{sec:related}

While still a relatively new and developing field, a growing collection of literature focuses on the privacy and security of people with disabilities using web services. 

\subsection{Differing Tool Usage Perceptions: Web Services}
Both on and offline, the general population and disabled population have vastly different needs and abilities. As technology advances, many adults increasingly use online services such as banking, social media, email, and healthcare~\cite{gitlow2014technology,das2017celebrities,mitzner2010older,walsh2021my,marston2016technology,das2021does,dev2018privacy}. As a result of this increase in technology use, many of these users have privacy and security concerns related to web services and how their data is being used~\cite{awotunde2021privacy,mccole2010trust,markert2023transcontinental}. While these web services can benefit users greatly, researchers such as Mentis et al. have found that they also create various privacy and security risks for vulnerable populations. In addition, many adults who use these services have mild cognitive impairment and other disabilities that make it difficult to understand the implications of sharing personal information online, the importance of password management, and recognizing scams~\cite{mentis2019upside,shrestha2022secureld,das2022sok,das2020non,das2020don,das2019don}. While these web services should make technology more accessible to all users, our SOK demonstrates that we need to perform an in-depth study to understand the needs of understudied populations.

\subsection{Privacy and Security Concerns}
When trying to understand more about how tool usage differs amongst these populations, the topic of authentication and CAPTCHA completion was at the forefront of six~\cite{helkala2012disabilities,shirali2009spoken,fuglerud2011secure,yan2008usability,ma2013investigating,kumar2020age} research papers. Authentication protocols are a hallmark of online privacy and security~\cite{das2019evaluating,jones2021literature,duezguen2020towards,majumdar2021sok}, necessary for all users to complete to gain access to their accounts or personal information. However, some authentication methods, such as CAPTCHAs, can be difficult or impossible for disabled users to complete since they rely heavily on visual outputs~\cite{patrick2022understanding,zhang2022building,mcleod2022challenges,jensen2021multi}. Therefore, Fuglerud et al. proposed a talking mobile one-time-password client that would provide users with both auditory and visual outputs~\cite{fuglerud2011secure}. This tool creates an environment where various types of users can complete authentication mechanisms without being overlooked based on their needs or abilities. However, our research reveals a scarcity of authentication tools and designs tailored to address the requirements of disabled populations.

\section{Methods}
\label{sec:methods}
Through this study, we aim to answer the following research questions (RQs):
\begin{itemize}
    \item \textit{RQ1: What are the privacy and security concerns related to the disabled community when interacting with web services?}
     \item \textit{RQ2: How can CAPTCHAs/authentication be improved to protect the privacy and security of people with disabilities for online communication?}
    \item \textit{RQ3: How can universal design, design for privacy, and inclusive privacy and security be implemented in different web services?}
\end{itemize}

To answer these questions, our literature review included several steps: (i) database search, (ii) title screening, (iii) duplicate removal, (iv) abstract screening,(v) full-text screening, and (vi)thematic analysis. Papers were included if they meet the following criteria: (1) Published in a peer-reviewed publication, (2) Published in English, (3) Technology discussed focuses on privacy and/or security of web services, (4) Target population includes a significant portion of individuals with disabilities. The exclusion criteria includes: (1) The technology discussed in the research work was not used primarily by people with disabilities, (2) The papers did not include a direct discussion of the privacy and security of users with disabilities for web services, (3) The paper was an abstract, poster, work-in-progress, or otherwise not a full paper, (4) The full-text of the papers were not available even after searching through multiple databases or after contacting the authors. Our methodology was adapted from prior works by Stowell et al.~\cite{stowell2018designing}, Das et al.~\cite{das2019all}, Tazi et al.~\cite{tazi2022sok,tazi2022sok1}, Noah and Das~\cite{noah2021exploring}, and Shrestha et al.~\cite{shrestha2022sok,shrestha2022exploring}.
\subsection{Database Search and Title Screening}
We conducted our search by exploring five digital databases, namely:IEEE Xplore~\footnote{https://ieeexplore.ieee.org/Xplore}, SSRN~\footnote{https://www.ssrn.com}, Google Scholar~\footnote{https://scholar.google.com/}, Science Direct~\footnote{https://www.sciencedirect.com/}, and ACM Digital Library~\footnote{https://dl.acm.org/}. The data collection spanned from May to July 2021 and included any paper published before July 2021.

We collected $14$ papers from IEEE Xplore, $3$ papers from SSRN, $1000$ papers from Google Scholar, $991$ papers from Science Direct, and $344$ papers from ACM Digital Library. The keyword search for IEEE Xplore, SSRN, and Science Direct was "disability + privacy + security," and the "research articles" filter was applied. For ACM Digital Library, the keyword search used was "disability" AND "privacy," AND "security" with the "full text" filter applied. We used the Publish or Perish~\cite{harzing2010publish} software to review Google Scholar articles. The keyword search used in Publish or Perish was "privacy and security" + "disabled people." This search was limited to $1000$ results by the software. We reviewed a total of $2,352$ article titles from all five databases. A paper was at this point deemed pertinent if the title discussed web services for people with disabilities, including those with specific impairments like visual, hearing, or motor impairments. Additionally, the title was required to describe a study investigating privacy and security concerns of using web services for the disabled population or the usage of web services in general about privacy or security. A paper was also only considered if it met the inclusion requirements. After duplicate removal, our corpus consisted of $138$ articles.

\subsection{Abstract and Full Text Screening }
We manually reviewed the abstracts of all $138$ papers in the research collection for relevance to our RQs. 

We removed $27$ papers during abstract screening, leaving $111$ papers for full-text screening. On these $111$ papers, we conducted a full-text screening where we reviewed the methods, findings, analysis, and discussions. 

After the full-text screening, $63$ relevant papers remained for the detailed thematic analysis.

\subsection{Data Extraction and Thematic Analysis}
For all $63$ papers remaining in our corpus, we extracted quantitative and qualitative findings to assess the web services' privacy and security perspectives on the disabled population-focused research conducted by prior studies. The extracted data included population samples, user experience, study design characteristics, and type of technology used (web services for our research). The results, discussion, and conclusion data from each paper were analyzed and coded according to themes identified by the first and third authors. The inter-coder reliability score for the coding was $89.4\%$. In places where the two authors could not agree, the second author decided. A random sample of $12$ papers was taken where the abstracts, methods, results, and discussions were reviewed. This resulted in themes such as:
\begin{itemize}
    \item Type of disability: visual impairments, Down Syndrome, cognitive disabilities
    \item Type of participant: some studies include both disabled and non-disabled people, while other studies include only disabled people
    \item Difficulty using authentication interfaces
    \item CAPTCHA completion can be hard or impossible for those who are blind, have low vision, or have a learning disability (dyslexia, ADHD.)
\end{itemize}

The remaining papers were then evaluated by going through each and generating a complete codebook. This process yielded a codebook that consists of $33$ overarching codes, which were themed into seven overarching themes including,\lq\lq~Authentication Interface Issues~\rq\rq, \lq\lq~Privacy Concerns as Reasons for Non-Use~\rq\rq,\lq\lq~Critical Data Access~\rq\rq,\lq\lq~Online Vulnerability~\rq\rq,\lq\lq~Solutions to authentica~\rq\rq,\lq\lq~Universal Design~\rq\rq and\lq\lq~Usability of Security Tools and Protocols~\rq\rq.

\section{Findings and Discussions}
\label{sec:findings}
In this section, we report on our findings while addressing the RQs mentioned in the previous section. 

\subsection{RQ1: Privacy and Security Concerns of Disabled People for Web Services}
Our first research question addresses the privacy and security concerns of people with disabilities when interacting with web services. We addressed this RQ by analyzing the different papers within the themes related to this specific research question which are four, namely:\lq\lq~Authentication Interface Issues~\rq\rq, \lq\lq~Privacy Concerns as Reasons for Non-Use~\rq\rq,\lq\lq~Critical Data Access~\rq\rq,\lq\lq~Online Vulnerability~\rq\rq. Table~\ref{tab:RQ1} provides the snapshot of the distribution of the papers which cater to RQ1. In the following subsections, we will provide more details about these themes.

\begin{table}[ht]
\begin{center}
\begin{tabular}{ |p{3cm}|p{4cm}| } 
 \hline
 Themes & Number of Papers\\
 \hline \hline
 Authentication Interface Issues & $4$ (6.35\%)~\cite{helkala2012disabilities,yan2008usability,ma2013investigating,bayor2018characterizing} \\ 
 Privacy Concerns as Reasons for Non-Use & $27$ (42.86\%) ~\cite{kordzadeh2016antecedents,ermakova2015antecedents,roberts2021evaluating,lafky2011personal,yao2012adoption,maqbool2021challenges,liang2017understanding,kaplan2019alzheimer,lafky2008prospective,ziefle2011medical,tantidham2019emergency,el2013trust,novitzky2015review,vasco2020internet,hadian2017efficient,pacheco2016secure,rahman2018blockchain,ramli2014privacy,wang2012intelligent,ifrim2017art,dang2019survey,hussain2021security,anisha2020automated,fosch2021cybersecurity,mohan2016security,kumar2020age,ahmed2020threats} \\ 
 Critical Data Access & $7$ (11.11\%)~\cite{solanas2013m,beach2009disability,hamidi2020using,singh2007password,hamidi2018should,grunwel2016delegation,distler2020acceptable} \\
 Online Vulnerability & $14$ (22.22\%)~\cite{onyeaka2020understanding,ahmed2015privacy,lazar2011understanding,macmillan2020autistic,jattamart2020perspectives,hersh2017mobile,scanlan2021reassessing,mentis2019upside,munozproviding,giannoumis2018accessibility,hersh2018mobility,king2019becoming,kim2019conceptualizing,chalghoumi2019information} \\
 \hline
\end{tabular}
\caption{The Distribution of Papers Across Themes Answering the RQ1}
    \label{tab:RQ1}
\end{center}
\end{table}

\subsubsection{Authentication Interface Issues}
Authentication is a basis of security standards and protocols for web services. While CAPTCHA completion and authentication steps are often easy for non-disabled users, the disabled population faces countless difficulties accessing their online services. While analyzing papers on security concerns for people with disabilities, we found that issues with authentication interfaces were a common theme discussed. We found underlying sub-themes, such as difficulty using authentication due to technical hindrances and how each disability can affect a user's capability to complete authentication mechanisms. Four papers from the $63$ in our corpus ~\cite{helkala2012disabilities,yan2008usability,ma2013investigating,bayor2018characterizing} relating to this category. One such paper discusses the success of CAPTCHA completion depending on the disabilities; for most non-disabled users, CAPTCHA completion and other forms of authentication are an almost unnoticeable part of using web services. 

However, users with any level of disability or impairment can find these same tasks to be difficult or impossible, as Helkala explains~\cite{helkala2012disabilities}. Through their work, Helkala explores how users with vastly different disabilities like Parkinson's disease, dyslexia, vision impairment, and upper extremity disabilities all experience different issues with CAPTCHA completion based on their abilities. In addition, this research raises important questions about how current authentication methods, such as static PIN codes, textual passwords, and one-time codes, can be altered better to fit different populations' needs and abilities.
Another equally important code within this theme is the difficulty of using authentication due to technical hindrances; these difficulties discussed were at the conceptual and adoption levels. This was detailed by Bayor et al. in their research analyzing interest in using social media amongst users with intellectual disabilities. Their findings suggest that a lack of accessible authentication methods for disabled users often hinders this interest. The authors also note that voice search, auto-login, and password retrieval protocols could be already-existing solutions for this user population~\cite{bayor2018characterizing}

\subsubsection{Privacy Concerns as Reasons for Non-Use}
In reviewing research papers on the privacy and security concerns of the disabled population when using web services, we found that an overwhelming majority of users cited privacy concerns as reasons for non-use. Every user wants their account and data to be protected from social media sites to healthcare technology. Some of the most prevalent sub-themes related to non-use were found in connection to medical technology in smart homes and concerns about health information technology used frequently by people with disabilities. If a user feels that their health information needs to be adequately protected, it was found that they often choose not to use the service at all. There are $27$ papers related to this theme, as detailed in table~\ref{tab:RQ1}. One such paper analyzes the privacy and security concerns of disabled people regarding medical technology used in smart homes.

Ziefle et al. researched the attitudes of disabled users towards a video-based monitoring system in the smart home environments of elderly or disabled people. They found that users would only feel comfortable with this system in their homes if strict privacy protocols were followed, including anonymity in transferring medical data, password protection, discretion, and avoidance of stigmatization~\cite{ziefle2011medical}. Furthermore, many health information technologies are becoming popular amongst users, especially smartphone apps and websites that access medical data. Onyeaka et al. discuss how it may be difficult for some user populations, such as those with disabilities or mental health conditions, to use these smartphone apps and websites. The researchers found that many users with disabilities would withhold crucial medical information from their healthcare providers because of privacy and security concerns about how their data was being used by the healthcare apps and websites~\cite{onyeaka2020understanding}. Concerns exist that these privacy and security issues could lead to further stigmatization and non-use by the disabled population.

\subsubsection{Critical Data Access}
We classified papers within \lq\lq~Critical Data Access\rq\rq~ if they discuss data sharing, specifically medical data, and the privacy and security concerns of disabled people over their critical data. Through these papers, we determine that users have privacy and security concerns related to sharing personal health records with caretakers, healthcare providers, insurance companies, researchers, and governments. In particular, many people with disabilities feel there are privacy trade-offs in emergency situations when they do not have control over who has access to their personal medical data. Seven papers from our corpus were included in this theme~\cite{solanas2013m,beach2009disability,hamidi2020using,singh2007password,hamidi2018should,grunwel2016delegation,distler2020acceptable}. One of these papers; Beach et al. discuss how technology aimed at enhancing independent living for people with disabilities is a growing field. However, there are still a lot of privacy and security concerns to consider. This is particularly relevant because the researchers found that users with disabilities are significantly more accepting of the sharing and recording personal medical information than non-disabled people~\cite{beach2009disability}. This raises concerns about how disabled people are more at risk of privacy and security failures than their non-disabled counterparts. On the other hand, Solanas et al. propose m-Carer, a smart mobile device that monitors patients' movements. The researchers hope to provide a way to track and find disabled users who become lost, disoriented, or need emergency medical attention~\cite{solanas2013m}. Although this new technology could help users in emergencies, it raises concerns about patient privacy invasions and how the tracking data is stored and transmitted.

\subsubsection{Online Vulnerability}
we classified papers that examine online vulnerabilities, particularly those that affect individuals with disabilities, as \lq\lq~Online Vulnerability\rq\rq. More than $22\%$ of the papers in our corpus fall under this theme, making it a prevalent one. ~\cite{onyeaka2020understanding,ahmed2015privacy,lazar2011understanding,macmillan2020autistic,jattamart2020perspectives,hersh2017mobile,scanlan2021reassessing,mentis2019upside,munozproviding,giannoumis2018accessibility,hersh2018mobility,king2019becoming,kim2019conceptualizing,chalghoumi2019information}. Many disabled users are unaware of the ever-changing nature of online privacy and security issues, and must rely on the assistance of a caregiver or family member to safeguard themselves. This raises concerns about the trade-offs between autonomy and privacy when disabled people use digital services. According to Chalghoumi et al., many disabled users are unaware of technology and web services' privacy and security concerns. The researchers found that the opinions of caregivers and family members of the disabled participant were significantly influential on the user's behavior toward online privacy~\cite{chalghoumi2019information}. This raises questions regarding how much of a disabled user's web services experience can be autonomous if caretakers substantially impact them.

\subsection{RQ2: Improving CAPTCHA/authentication}
The second RQ focuses on how CAPTCHAs/authentication can be improved to protect the privacy and security of people with disabilities when using web services. Some disabled users can find authentication completion impossible and are consequently unable to access their accounts. Six papers~\cite{shirali2009spoken,brown2010using,barbosa2016unipass} from our corpus focus on solutions to improving authentication and CAPTCHAs. Table~\ref{tab:RQ2} provides the snapshot of the distribution of these papers.

\begin{table}[ht]
\begin{center}
\begin{tabular}{ |p{3cm}|p{4cm}| } 
 \hline
 Theme & Number of Papers\\
 \hline \hline
 Solutions to authentication/CAPTCHA Issues & $3$ (4.76\%) ~\cite{shirali2009spoken,barbosa2016unipass,brown2010using} \\ 
 \hline
\end{tabular}
\caption{The Distribution of Papers Across Themes Answering the RQ2}
    \label{tab:RQ2}
\end{center}
\end{table}
Some papers relating to this theme provided the solution to authentication problems; one such solution is using passtones instead of passwords, as researched by Brown and Doswell. Rather than remembering alphanumeric sequences, Brown and Doswell propose a password alternative where users would remember a sequence of sounds~\cite{brown2010using}. The researchers explain how this tool has already been implemented using photos, but using auditory passwords would improve the experience of users with visual disabilities. While explicitly a solution for visually impaired users, this solution could be widely implemented and used by people of all different needs and abilities. Similarly, accessible password managers are another solution to issues with authentication that many users face. Barbosa et al. describe their implementation of UniPass, an accessible password manager for visually impaired users on a smart device. This tool includes features such as reading prompts and messages aloud, buttons and other graphical elements are avoided, and the device vibrates to signify the need for user input~\cite{barbosa2016unipass}. The researchers found that password managers are a promising solution for the difficulties visually impaired users face with authentication mechanisms. A different way to enhance the authentication experience of disabled users when interacting with web services is Spoken CAPTCHA. Shirali-Shahreza et al. discuss how most CAPTCHA methods currently only use visual patterns, making it impossible for blind users to complete them. The researchers propose a new CAPTCHA method, Spoken CAPTCHA, where users would hear a short sound clip asking them to say a word. The user will then respond in a speech file that can be checked not to be computer generated~\cite{shirali2009spoken}. This solution focuses on the visually impaired population and provides a way to improve authentication methods for all types of users.

\subsection{RQ3: Universal Design, Design for Privacy, and Inclusive Privacy and Security in Web Services}
The third RQ focuses on how universal design, design for privacy, and inclusive privacy and security can be implemented in different web services. These inclusive concepts provide design tools and protocols to make web services more accessible for various user populations, regardless of needs and abilities. We have gleaned two themes pertaining to this research question,\lq\lq~Universal Design~\rq\rq and\lq\lq~Usability of Security Tools and Protocols~\rq\rq. Table~\ref{tab:RQ3} provides the snapshot of the distribution of the papers which caters to the RQ3.

\begin{table}[ht]
\begin{center}
\begin{tabular}{ |p{3cm}|p{4cm}| } 
 \hline
 Theme & Number of Papers\\
 \hline \hline
 Universal Design & $6$ (9.53\%) ~\cite{vales2008ucare,wang2018inclusive,wang2017third,o2017privacy,mcrae2020privacy,medley1998ethical} \\ 
 Usability of Security Tools and Protocols & $2$ (3.17\%) ~\cite{han2018proximity,fuglerud2011secure} \\ 
 \hline
\end{tabular}
 \caption{The Distribution of Papers Across Themes Answering the RQ3}
    \label{tab:RQ3}
\end{center}
\end{table}

\subsubsection{Universal Design}
The Universal Design concept describes how the design of all products and environments should be usable by all people without the need for adaptation or specialized design. Inclusive privacy and security and privacy by design are closely related to the overarching theme of universal design. Six papers~~\cite{vales2008ucare,wang2018inclusive,wang2017third,o2017privacy,mcrae2020privacy,medley1998ethical} were included in this theme. These papers discuss the current privacy and security protocols that are most widely used and why they do not consider the needs and abilities of under-served populations such as children, older adults, people with disabilities, and people from non-Western populations. Wang et al. discuss the implementation of inclusive privacy and security tools, and protocols would prioritize the design of mechanisms that are inclusive to people with various characteristics, abilities, needs, and values~\cite{wang2018inclusive}. Similarly, we considered papers on privacy by design and how designers and technologies must put inclusive privacy and security tools/protocols at the forefront of their design. One of the most practical ways these designers can implement privacy by design is to increase digital citizen awareness surrounding consent for data processing and usage. O'Connor et al. discuss how users must have the information they need to make informed decisions about how their data is being used~\cite{o2017privacy}.

\subsubsection{Usability of Security Tools and Protocols}
The usability and accessibility of security tools and protocols are essential to the overarching theme of universal design. While the previous theme describes the theory of universal design, this theme explores implementations of the theory. The two papers related to this theme~\cite{fuglerud2011secure,han2018proximity} present inclusive password management and two-factor authentication solutions for various user populations across two related papers.
Password protection is a hallmark of online security tools and protocols. However, complicated authentication procedures to access web services can be cumbersome, especially for people with disabilities or the elderly. According to Fuglerud et al., a secure and accessible multi-modal authentication method using a one-time password client could solve this problem. Users with impairments affecting their ability to complete authentication steps now have access to auditory and visual outputs from the password client~\cite{fuglerud2011secure}. This allows all users equal access to password management tools and protocols.
The second paper by Han et al. describes how current 2FA solutions all require some form of user effort, with can negatively impact the experience of disabled users or the elderly. Therefore, the researchers propose a new type of mobile 2FA, Proximity-Proof, that does not require user interactions and defends against the powerful man-in-the-middle attack~\cite{han2018proximity}. According to the authors, Proximity-Proof is as secure as other 2FA methods and provides innovative ways for 2FA techniques to become more usable and accessible for all users.

\section{Future Work and Limitation}
\label{sec:future}

In this paper, we conducted a systematic analysis to evaluate the research articles and peer-reviewed papers published in the field of security and privacy of web services for the disabled population. We collected papers from five digital databases and limited the papers to ones available in English. As such we might have missed papers not available in these databases. However, our extensive literature review provides a detailed overview of the current research on security and privacy of web services for the disabled population. And while this gives a broad understanding of the current research and methods used, there is limited in-depth research on individual user groups within the disabled population. For example, five of the six papers relating to solutions for authentication issues were only solutions for visually impaired users. Future analyses of privacy and security concerns of the disabled population can provide valuable research into more specific subsections of the population, such as those with cognitive disabilities, mental illnesses, and different types of physical impairments.

\section{Conclusion}
\label{sec:conclusion}
For many disabled users, information technology and web services can be a way to enhance their autonomy and discover new interests or communities. However, disability can make the internet a challenging place, seeing as many disabled people have trouble writing, reading, and comprehending text information, making it hard for them to understand and use basic security and privacy measures such as passwords and passwords CAPTCHAs. To that regard, we conducted a systematic literature review on $63$ papers focused on the privacy and security of web services for the disabled population. Our findings reveal valuable solutions to privacy and security concerns of the disabled population, focused on universal design and inclusive privacy and security methods. Universal design, in particular, provides a way to create inclusive, accessible, and usable tools and protocols to protect the privacy and security of both the disabled and general populations online. These solutions would address issues such as authentication improvement, critical data access, online vulnerability, and usability of tools and protocols. However, our findings reveal gaps in the current research, such as a lack of implementation of these universal design methods and how solutions must focus on more subsections of the disabled population.

\section{Acknowledgement}
We would like to thank the Inclusive Security and Privacy focused Innovative Research in Information Technology (InSPIRIT) Laboratory at the University of Denver. This research has been funded by the Faculty Research Fund (FRF) at the University of Denver. Any opinions, findings, conclusions, or recommendations expressed in this material are solely those of the authors and not of the organization or the funding agency.

\bibliographystyle{IEEEtran}

\bibliography{main}

\end{document}